\documentclass[twocolumn,aps,showpacs]{revtex4}
\usepackage[dvips]{graphics,color}
\usepackage{dcolumn}
\usepackage{bm}
\usepackage{epsfig}
\usepackage{graphicx}

\newcommand{\beq}{\begin{equation}}
\newcommand{\eeq}{\end{equation}}
\newcommand{\bd}[1]{ \mbox{\boldmath $#1$}}

\begin{document}
\def\ii{\'\i}

\title{
Modelling Pentaquark and Heptaquark States
}

\author{
M. Nu\~nez V.,
S. Lerma H. and
P. O. Hess,}
\affiliation{Instituto de Ciencias Nucleares, Universidad
Nacional Aut\'onoma de M\'exico \\
Apdo. Postal 70-543, M\'exico 04510 D.F.}
\author{S. Jesgarz}
\affiliation{Instituto de Fisica, Unversidade de S\~ao Paulo,
CP 66318, S\~ao Paulo, 05315-970, SP, Brasil}
\author{O. Civitarese and
M. Reboiro}
\affiliation{Departamento de F\'{\i}sica, Universidad Nacional de La Plata, \\
c.c. 67 1900, La Plata, Argentina. }

\begin{abstract}
{A schematic model for hadronic states, based on constituent
quarks and antiquarks and gluon pairs, is discussed. The
phenomenological interaction between quarks and gluons is QCD
motivated. The obtained hadronic spectrum leads to the
identification of nucleon and $\Delta$ resonances and to
pentaquark and heptaquark states. The predicted lowest pentaquark
state ($J^{\pi} = \frac{1}{2}^-$) lies at the energy of 1.5 GeV
and it is associated to the observed $\Theta^+(1540)$ state. For
heptaquarks ($J^{\pi} = \frac{1}{2}^+, \frac{3}{2}^+$) the model
predicts the lowest state at 2.5 GeV. } 
\end{abstract}

\maketitle


In a series of previous publications
\cite{paperI,paperII,paperIII} a schematic model for QCD was
developed. The model was used to test the meson spectrum of QCD.
In spite of its schematic nature the model seems to contain the
relevant degrees of freedom, as it was shown in the comparison
between calculated and experimental meson spectra \cite{paperII}.
This letter is devoted to the extension of the model to
accommodate baryonic features. Particularly, we shall concentrate
on the appearance of exotic baryonic states, like pentaquark and
heptaquark states \cite{exp1,exp11,exp2,exp22,exp3}.

The essentials of the model were discussed in detail in Ref.
\cite{paperII}. It consists of two fermionic levels in the quark
($q$) and antiquark ($\bar{q}$) sector and a gluonic ($g^2$)
state
containing pairs of gluons.
These are the elementary degrees of freedom of the model. The
interaction among these degrees of freedom is described by
excitations of pairs of quarks and antiquarks mediated by the
exchange of pairs of gluons. The pairs of quarks are classified in
a flavor-spin coupling scheme. The pairs of gluons are kept in the
angular momentum ($J$), parity ($\pi$) and charge conjugation
($C$) state $J^{\pi C} = 0^{++}$. The strength of various channels
of the interaction, as well as the constituent masses, are taken
from a phenomenological analysis. The model describes meson
($(q\bar{q})^n (g^2)^m$) states and baryonic ($q^3(q\bar{q})^n
(g^2)^m$) states. Among these states we focus on $q^3(q\bar{q})$
states (pentaquarks) and $q^3(q\bar{q})^2$ states (heptaquarks),
where the configurations indicated represent the leading terms in
an expansion over many quark-antiquark and gluon states.
The basis states are classified using group theoretical methods
\cite{paperII}. The interaction of quark-antiquark pairs with
gluon pairs is particle non-conserving.

The above described model belongs to a class of exactly solvable
models of coupled fermion and boson systems
\cite{lipkin,schutte,pittel,ocmr}. Alternative descriptions of
pentaquark states were proposed in Ref. \cite{bijker}, enforcing
particle number conservation.

In what follows we shall classify the basis states and solve the
Hamiltonian in the framework of the boson expansion method
\cite{klein,chh}. Finally, we shall compare the results of the
calculations with recently published experimental data
\cite{exp1,exp11,exp2,exp22}


The model Hamiltonian is written

\begin{eqnarray}
\bd{H} & = & 2\omega_f \bd{n_f} + \omega_b \bd{n_b} \nonumber \\
&+&\sum_{\lambda S} V_{\lambda S} \left\{ \left[ (\bd{b}_{\lambda
S}^\dagger )^2 + 2\bd{b}_{\lambda S}^\dagger \bd{b}_{\lambda S} +
(\bd{b}_{\lambda S})^2 \right] (1-\frac{\bd{n_f}}{2\Omega})\bd{b}
\right.
\nonumber \\
&+& \left. \bd{b}^\dagger (1-\frac{\bd{n_f}}{2\Omega})
 \left[ (\bd{b}_{\lambda S}^\dagger )^2 +
2\bd{b}_{\lambda S}^\dagger \bd{b}_{\lambda S}+ (\bd{b}_{\lambda S})^2
\right] \right\}       \nonumber \\
&+ & \bd{n}_{(0,1)0} \left( D_1 \bd{n_b} + D_2(\bd{b}^\dagger +
\bd{b}) \right)\nonumber \\& +& \bd{n}_{(2,0)1} \left( E_1
\bd{n_b} + E_2(\bd{b}^\dagger + \bd{b}) \right) ~~~.
\label{hamiltonian}
\end{eqnarray}
The distance between the fermion levels is $2\omega_f$=0.66 GeV,
$\omega_b$=1.6 GeV is the energy of the glue ball, $\bd{n_f}$ and
$\bd{n_b}$ are the number operators for fermion and gluon pairs,
respectively, $V_{\lambda S}$ is the strength of the interaction
in the flavor$(\lambda$) and spin ($S$) channel. The actual values
$\lambda$ = 0, 1 refer to flavor (0,0) and (1,1) configurations,
while the spin channel is $S$=0 or 1. The adopted values are:
$V_{00}$=0.0337~GeV, $V_{01}$=0.0422~GeV, $V_{10}$=0.1573~GeV, and
$V_{11}$=0.0177~GeV \cite{paperII} . The operators
$\bd{b}_{\lambda S}^\dagger$ and $\bd{b}_{\lambda S}$ are boson
images of quark-antiquark pairs \cite{paperII}. The products which
appear inside brackets in (\ref{hamiltonian}) are scalar products.
The factor $(1-\frac{\bd{n_f}}{2\Omega})$ results from the boson
mapping \cite{paperII}.
The mapping is exact for the channel $[\lambda , S]=[0,0]$ and
simulates the effect of the boson mapping for the other channels.
The operator $\bd{b}^\dagger$ ($\bd{b}$) creates (annihilates)
gluon pairs with spin-color zero, and
$\bd{n}_{(\lambda_0,\mu_0)S_0}$ is the number operator of
di-quarks coupled to flavor-spin $(\lambda_0,\mu_0)S_0$. The
parameters $D_{1(2)}$ and $E_{1(2)}$ are adjusted to the nucleon
and $\Delta$ resonances. The corresponding terms describe the
interaction between valence quarks and glueballs. The Hamiltonian
(\ref{hamiltonian}) does not contain terms which distinguish
between states with different hypercharge and isospin. It does not
contain flavor mixing terms, either. Therefore, the predicted
states have to be corrected in the way described in \cite{gursey}
to allow a comparison with data. The adopted values of $D_{1(2)}$
and $E_{1(2)}$ are: $D_1=$-1.442GeV, $D_2$=-0.4388GeV,
$E_1$=-1.1873GeV and $E_2$=-0.3622GeV.
The Hamiltonian contains all relevant degrees of freedom requested by QCD.


The complete classification of quark-antiquark configurations was given
in Ref. \cite{paperII}.

The unperturbed ground state is composed by 18 quarks occupying
the lowest fermionic level. The baryonic states are described by
three quarks in the upper fermionic level to which we add
$(q\bar{q})^n$ states. The group chain which describes these
states is

\begin{eqnarray}
[1^N] &  [h]=[h_1h_2h_3] & [h^T] \nonumber \\
U(4\Omega )  & \supset  U(\frac{\Omega}{3}) \otimes & U(12)
\label{group1}
\end{eqnarray}
where $\Omega$=9 accounts for three
color and three flavor degrees of freedom. The irreducible
representation (irrep) of $U(4\Omega )$ is completely
antisymmetric, and $[h^T]$ is the transposed Young diagram of
$[h]$ \cite{hamermesh}. For $N$ particles, and due to the
antisymmetric irrep $[1^N]$ of $U(4\Omega)$, the irreps of
$U(\Omega /3)$ and $U(12)$ are complementary and the irrep of
$U(\Omega /3)$ is the color group,
which is reduced to $SU_C(3)$ with the color irrep $(\lambda_C,\mu_C)$.
The $U(12)$ group is further reduced to

\begin{eqnarray}
U(12) \supset & U_f(3) \otimes U(4) & \supset SU_f(3) \otimes SU_S(2)
\nonumber \\
&~~~~~ [p_1p_2p_3p_4] & (\lambda_f,\mu_f) ~~~~ S,~M ~~~,
\label{u12}
\end{eqnarray}
where $(\lambda_f,\mu_f)$ is the flavor irrep and $[p_1p_2p_3p_4]$
denotes the possible $U(4)$ irreps. The group reduction is
done using the methods exposed in Ref. \cite{ramon}.
The basis is spanned by the
states
\begin{eqnarray}
|N, [p_1p_2p_3p_4] (\lambda_C ,\mu_C), \rho_f (\lambda_f,\mu_f) Y T T_z, \rho_S S M> ~~~,
\label{state}
\end{eqnarray}
where $N$ is the number of particles, $Y$ is the hypercharge and
($T$,$T_z$) denotes the isospin and its third component, $\rho_f$
and $\rho_S$ are the multiplicities of the flavor and spin
representations. The color labels $(\lambda_C,\mu_C)$ are related
to the $h_i$ via $\lambda_C = h_1-h_2$ and $\mu_C = h_2-h_3$. To
obtain the values of $h_i$ one has to consider all possible
partitions of $N=h_1+h_2+h_3$, which fixes the color. For
colorless states we have $h_1=h_2=h_3=h$. Each partition of $N$
appears only once. The irrep $[hhh]$ of $U(\frac{\Omega}{3})=U(3)$
fixes the irrep of $U(12)$, as indicated in (\ref{group1}). For
$\Omega = 9$ and color (0,0) the irrep of $U(12)$ is given by
$[3^6 0^6]$ for mesons, and by $[3^7 0^5]$ for baryons.
As an example,
Table \ref{table1} shows the relevant irreps for mesonic states.
(More details are given in Ref. \cite{long}).
\begin{center}
\begin{table}[th]
\begin{center}
\begin{tabular}{|l|l|l|l|l|l|l|l|l|}
\hline
$SU_f(3)$ & $U(4)$ & $[q_1 q_2]$ & $n_q$ & $S_q$ & $[\bar{q}_1 \bar{q}_2]$ & $n_{\bar{q}}$
& $S_{\bar{q}}$ & $S$ \\
\hline
(0,0), (1,1), (2,2) & $[8811]$ & $[11]$ & 2 & 0 & $[88]$ & 2 & 0 & 0 \\
(1,1), (3,0), (0,3) & $[9711]$ & $[11]$ & 2 & 0 & $[97]$ & 2 & 1 & 1 \\
(1,1), (3,0), (0,3) & $[8820]$ & $[20]$ & 2 & 1 & $[88]$ & 2 & 0 & 1 \\
(0,0), (1,1), (2,2) & $[9720]$ & $[20]$ & 2 & 1 & $[97]$ & 2 & 1 & 0, 1, 2 \\
(1,1) & $[9810]$ & $[10]$ & 1 & $\frac{1}{2}$ & $[98]$ & 1 & $\frac{1}{2}$ & 0, 1 \\
(1,1) & $[9810]$ & $[11]$ & 2 & 0 & $[97]$ & 2 & 1 & 1 \\
(1,1) & $[9810]$ & $[11]$ & 2 & 0 & $[88]$ & 2 & 0 & 0 \\
(1,1) & $[9810]$ & $[20]$ & 2 & 1 & $[97]$ & 2 & 1 & 0, 1, 2 \\
(1,1) & $[9810]$ & $[20]$ & 2 & 1 & $[88]$ & 2 & 0 & 1 \\
(0,0) & $[9900]$ & $[00]$ & 0 & 0 & $[99]$ & 0 & 0 & 0 \\
(0,0) & $[9900]$ & $[10]$ & 1 & $\frac{1}{2}$ & $[98]$ & 1 & $\frac{1}{2}$ & 0, 1 \\
(0,0) & $[9900]$ & $[20]$ & 2 & 1 & $[97]$ & 2 & 1 & 0, 1, 2 \\
\hline
\end{tabular}
\end{center}
\caption{ Flavor irreps coupled to the quark-antiquark content of
some different $U(4)$ irreps. Shown are the irreps which contain,
at most, two quarks and two antiquarks. The number of quarks
(antiquarks) in a given configuration are denoted by $n_q$
($n_{\bar{q}}$). } \label{table1}
\end{table}
\end{center}
In the boson representation, the states are given by the direct
product of one-, three-, eight and 24-dimensional harmonic
oscillators \cite{paperII}. For each harmonic oscillator the basis
states are given by

\begin{eqnarray}
{\cal N}_{N_{\lambda S}\nu_{\lambda S}} (\bd{b}^{\dagger~2}_{\lambda S}
)^{\frac{N_{\lambda S}-\nu_{\lambda S}}{2}}
|\nu_{\lambda S}\alpha_{\lambda S} > ~~~,
\label{basis}
\end{eqnarray}
where $N_{\lambda S}$ is the number of bosons of type
$[\lambda ,S]$, $\nu_{\lambda S}$ is the corresponding seniority
and ${\cal N}_{N_{\lambda S}\nu_{\lambda S}}$ is a normalization constant.
The seniority is defined as the number of uncoupled bosons.
The quantity $\alpha_{\lambda S}$ represents the other quantum numbers
needed to specify a
particular harmonic oscillator.

\section*{a) Nucleon resonances}

The quality of the model predictions, concerning meson states, was
discussed in Ref. \cite{paperII}. Figure 1 shows the lowest
nucleon and $\Delta$ resonances predicted by the model. In the
same Figure are shown the calculated penta- and heptaquark low
lying states. For each state we indicate the spin, parity
($J^{\pi}$), and the quark and gluon content ($n_q+n_{\bar{q}}$,
$n_g$). The quantity $n_q+n_{\bar{q}}$ is the total number of
quarks and antiquarks, which is equal to the number of valence
quarks (0 for mesons, 3 baryons) plus the number of quarks and
antiquarks of the $q\bar{q}$-pairs, and $n_g$ gives the number of
gluons. As shown in the Figure, nucleonic states contain
on the average
about
half an additional quark-antiquark pair (equivalent to one
extra quark), and approximately 2.8 gluons. This implies a content
of 59$\%$ in the quark sector and of 41$\%$ in the gluon sector.
The theoretical Roper resonance
(first excited nucleon resonance)
lies near the experimental energy of 1.44~GeV. This is a nice
feature of the model, which is shared only with the constituent
quark model \cite{bijker-r}. The first negative parity state with
$J^{\pi}=\frac{1}{2}^-$ appears at 1.51~GeV, also in good
agreement with the data.
\begin{figure}[t]
\includegraphics[width=9.0cm]{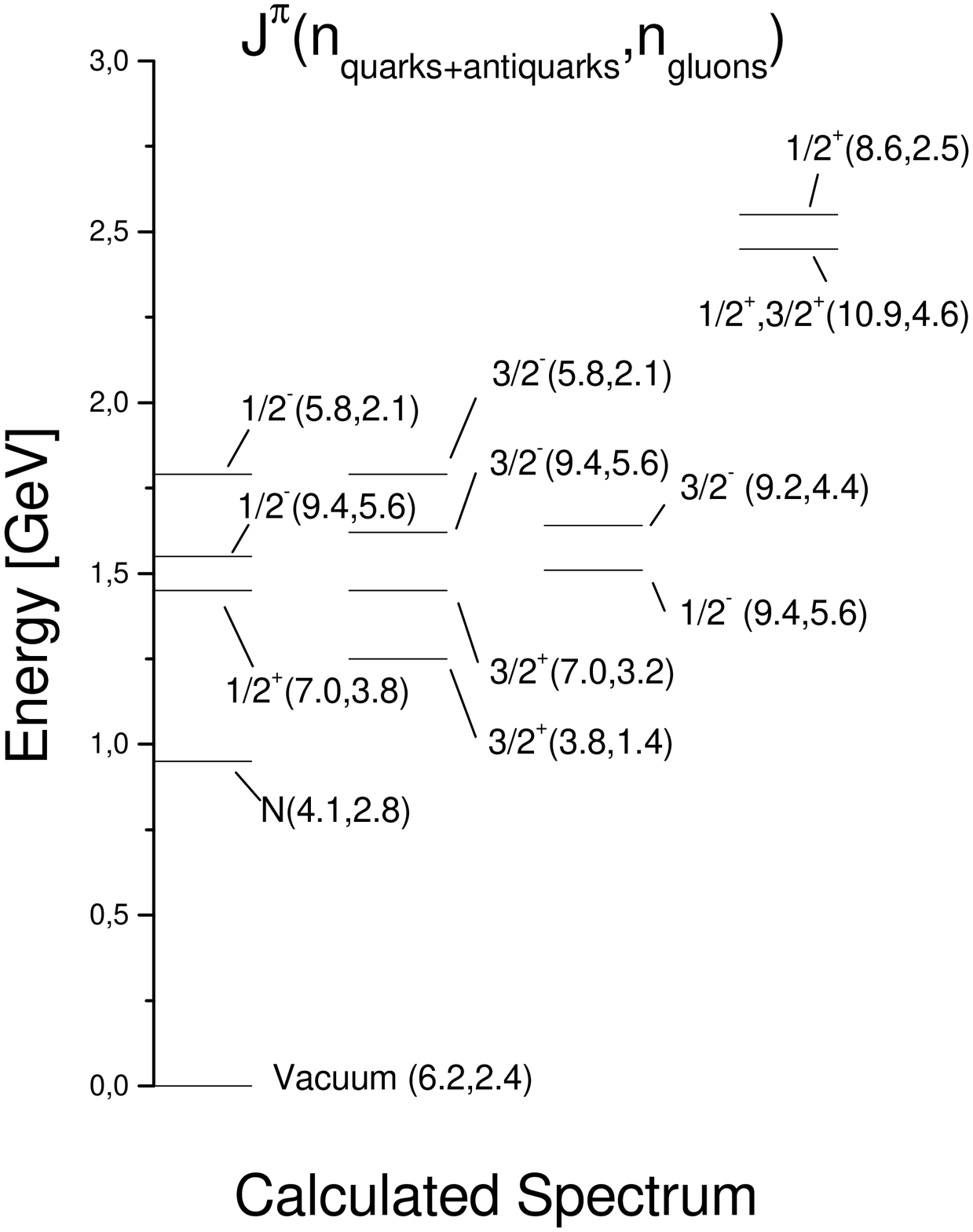}
\caption{ Nucleon resonances (first group of levels), $\Delta$
resonances (second group), pentaquarks (third group) and
heptaquarks (fourth group). On the right side of each level are
given the assigned spin and parity ($J^\pi$), and the total quark
and anti-quark
($n_q+n_{\bar{q}}$) and  gluon ($n_g$) contents (see the text)}
\label{fig-pref1}
\end{figure}

\section*{b) $\Delta$ resonances}

The simplest way to obtain a $\Delta$ resonance is to couple the
three valence quarks in the (3,0)$\frac{3}{2}$ configuration with
$q{\bar q}$ pairs in a (0,0)$J=0$ configuration. This scheme leads
to the $\Delta (1232)$ resonance. The quark-antiquark and gluon
content of the calculated $\Delta (1232)$ turns out to be lower
than that of the nucleon, while the structure of the state at
1.57GeV can be compared with the Roper resonance. Concerning
negative parity states, Fig. 1 shows a $\frac{3}{2}^-$ state at
1.79~GeV. $\Delta$ resonances can also be obtained by coupling the
three valence quarks in the $(1,1)\frac{1}{2}$ configuration with
 $(1,1)S$ $q\bar{q}$ states. The lowest state of this type is at
1.51~GeV, and it should be compared with the experimental value
($\Delta$ resonance) at 1.62~GeV \cite{databook}.

\section*{c) Pentaquarks and heptaquarks}

In the present calculation the minimal representation of
pentaquark-like states includes the following configurations:
(0,0)$\frac{1}{2}^-$, (1,1)$\frac{1}{2}^-$, (3,0)$\frac{1}{2}^-$,
(0,3)$\frac{1}{2}^-$ and (2,2)$\frac{1}{2}^-$. Only the
(0,3)$\frac{1}{2}^-$
and (2,2)$\frac{1}{2}^-$ configurations
contains hypercharge and isospin combinations which can not be
obtained in a pure $q^3$ coupling scheme,
like $T=0$, $Y=2$ in (0,3) and $T=1$, $Y=2$ in (2,2).
Within the model, the
lowest pentaquark state $\Theta^+(1540)$ \cite{exp1,exp11,exp2,exp22,exp3} is
interpreted as a coupling of the three valence quarks in
(1,1)$\frac{1}{2}^+$ with the $q\bar{q}$ background in (1,1)$0^-$
to the final irrep (1,1)$\frac{1}{2}^-$. Thus, within our model,
the calculated pentaquark state at 1.51~GeV may correspond to the
observed $\Theta^+(1540)$ state \cite{exp1,exp11,exp2,exp22,exp3}.
Anther
predicted pentaquark
state is
shown in Figure. 1.

Within the model, the lowest
pentaquark has negative parity in accordance with QCD sum-rules
and lattice gauge
calculations \cite{sum1,sum2,lat1,lat2}.
If the orbital spin $L$ is included, pentaquark states with positive
parity may exist with $L$=1. However, these states include an orbital
excitation and should appear at higher energies.

The model contains heptaquarks, characterized by two $q\bar{q}$
pairs added to the three valence quarks. The lowest state has an
energy of approximately 2.5~GeV and it has a content of
3.9
$q\bar{q}$ pairs of the type
$(1,1)1^+$
coupled to the three
valence quarks in the configuration (1,1)$\frac{1}{2}^+$. This
coupling scheme yields three exotic flavor irreps:
(3,3)$\frac{1}{2}^+$,$\frac{3}{2}^+$,
(4,1)$\frac{1}{2}^+$,$\frac{3}{2}^+$ and
(1,4)$\frac{1}{2}^+$,$\frac{3}{2}^+$.
The lowest heptaquark state contains,
basically,
three ideal valence quarks,
four
$q{\bar q}$ pairs and
4.6
gluons. This implies a quark content of
70$\%$ and a gluon content
of 30$\%$.
The model predicts other heptaquark states at higher
energies, which are obtained by coupling the three ideal valence
quarks with the $(3,0)1^+$ and $(0,3)1^+$ $q\bar{q}$
configurations. This leads to exotic flavor irreps  like (4,1),
(1,4) and (3,3) with spin $\frac{1}{2}^+$ and $\frac{3}{2}^+$. The
coupling of the three valence quarks with a $q\bar{q}$ irrep (2,2)
$S$= 0, 1, 2 leads to exotic flavor irreps of the type (3,3),
(1,4) with spin-parity $\frac{1}{2}^+$, $\frac{3}{2}^+$ and
$\frac{5}{2}^+$ \cite{long}.

\section*{d) Higher multiquark states}

In this letter we do not go further into the discussion of all
possible states with the structure $q^3(q\bar{q})^{n_2}g^{n_3}$,
since the number of these configurations increases with the
energy. A more complete overview of these states will be
presented in Ref. \cite{long}
with its complete classification of states.

\vskip 0.5cm

To conclude, we have applied a schematic model based on QCD
degrees of freedom, to describe nucleon and $\Delta$ resonances
and more exotic penta- and heptaquark states. The basis states
were classified by applying group theoretical methods. The
Hamiltonian, used in the calculations, was tested to the mesonic
spectrum, nucleon and $\Delta$ resonances. After fixing the
parameters in this manner, we have investigated the appearance of
penta- and heptaquark states. The results of the calculations show
that the model predicts reasonably well the $\Theta^+$(1540)
resonance. The lowest pentaquark state is obtained at an energy of
approximately 1.5~GeV and it has $J^\pi$=$\frac{1}{2}^-$. The
lowest heptaquark state has an energy of approximately 2.5~GeV and
$J^\pi$ either $\frac{1}{2}^+$ or $\frac{3}{2}^+$.
In addition, other penta- and heptaquark
states are predicted to appear at higher energies.

The model allows for a complete classification of many
quark-antiquark and gluon systems. As we have shown, the exotic
configurations which appear in our classification scheme can not
be obtained in a simple constituent quark picture. The overall
agreement with the experimental data supports the claim about the
suitability of the procedure.

\section*{Acknowledgments}
This work obtained financial support from the DGAPA project IN119002
and from CONACyt, CONICET. SJ thanks the {\it German Academic Interchange Service}
(DAAD) for financial support.

\end{document}